\documentclass[12pt]{iopart}
\usepackage{graphicx}
\usepackage{indentfirst}

\begin{document}

\title[]{Singularity of classical and quantum correlations at critical
points of the Lipkin-Meshkov-Glick model in bipartition and tripartition of
spins}
\author{Xiu-xing Zhang$^{1,2}$ and Fu-li Li$^{1,\dag}$}
\address{$^{1}$MOE Key Laboratory for Nonequilibrium Synthesis and Modulation of
Condensed Matter, and Department of Applied Physics, Xi'an
Jiaotong
University, Xi'an 710049, China\\
$^{2}$Department of Physics, Weinan Normal University, Wei'nan
714000, China} \ead{flli@mail.xjtu.edu.cn}
\begin{abstract}
We study the classical correlation (CC) and quantum discord (QD) between two
spin subgroups of the Lipkin-Meshkov-Glick (LMG) model in both binary and
trinary decompositions of spins. In the case of bipartition, we find that
the classical correlations and all the quantum correlations including the
QD, the entanglement of formation (EoF) and the logarithmic negativity (LN)
are divergent in the same singular behavior at the critical point of the LMG
model. In the case of tripartition, however, the classical correlation is
still divergent but all the quantum correlation measures remain finite at
the critical point. The present result shows that the classical correlation
is very robust but the quantum correlation is much frangible to the
environment disturbance. The present result may also lead to the conjecture
that the classical correlation is responsible for the singularity behavior
of physics quantities at critical points of a many-body quantum system.
\end{abstract}
\pacs{03.67.Bg, 75.10.Jm, 03.65.Ud, 64.70.Tg} $^{\dag }$Author to
whom any correspondence should be addressed
 \maketitle
\newpage

\section{Introduction}

In a many-body quantum system, the interplay of various energies leads to
different phases. When one of the energies becomes dominant over all the
others by varying either adjustable interaction constants or applied
external fields in the Hamiltonian, the system undergoes a phase transition
and some observable display the singular behavior at a critical point. Since
this phase transition occurs at zero temperature and is induced purely by
quantum fluctuations in the system, in contrast to usual phase transitions
induced by the thermal fluctuation, it is called quantum phase transition
(QPT) \cite{1}. During the last decade, the QPT has attracted a lot of
attention and become an important research domain \cite{1}. On the other
hand, quantum many-body systems have genuinely "quantum" correlations or
entanglement in contrast to classical correlations \cite{2,3}. Therefore, it
becomes natural to connect QPTs to quantum entanglement. At present, many
measures for entanglement have been proposed such as the relative entropy
\cite{4}, the concurrence \cite{5}, the entanglement of formation \cite{6},
the logarithmic negativity \cite{7} and so on. As observables for
identifying QPTs, those quantities indeed display the singular behaviour at
zero temperatures \cite{10,11,LMG4}. When calculating the measures of
entanglement, one need to divide a system into several subsystems and then
investigate quantum correlations between the subsystems. For a bipartition,
two subsystems are of complementary parts one another and the entire system
is always in a pure state. For a more multi partition, however, any two
subsystems no longer forms a whole system and are in general in a mixed
state. In this case, the other parts play a role of environments to the two
subsystems under consideration and may strongly affect the critical behavior
of quantum correlations between the chosen subsystems. In fact, Osborne
\textit{et al.} \cite{chain1} investigated the two-spin entanglement in the
XY spin model and found that the entanglement remains finite and displays a
peak at the critical point. Vidal \textit{et al.} \cite{chain2} studied the
entanglement of a $L$ spins block and found that the entanglement entropy
displays a logarithmic divergence for large $L$. Recently, Werlang \textit{%
et al.} \cite{13} showed that the quantum discord (QD) \cite{QD1,QD2} and
the entanglement of formation (EoF) between nearest-neighbors spins in an
XXZ infinite spin chain at finite temperatures are no longer divergent but
become finite at the critical point.

The Lipkin-Meshkov-Glick model (LMG) is one of few solvable many-body
systems. In recent years, a lot of efforts have been devoted to the study of
quantum correlations such as the entanglement entropy, the concurrence and
the logarithmic negativity in the LMG model \cite%
{LMG1,LMG2,LMG3,LMG5,LMG6,LMG7,LMG8,LMG9,LMG10,LMG11,LMG13}. Latorre \textit{%
et al.} \cite{LMG3} investigated the entanglement entropy in the LMG model
and found that the entropy displays a singularity at the critical point.
Morrison \textit{et al.} \cite{LMG7} studied the dynamical QPTs with the
spin-spin entanglement in a dissipative LMG. Or\'{u}s \textit{\ et al.} \cite%
{LMG9} investigated the many-body entanglement in the LMG model and showed
that the critical scaling laws for the single-copy entanglement and the
global geometric entanglement are equivalent. Wichterich \textit{\ et al.}
\cite{LMG14} found that the logarithmic negativity of two macroscopic sets
of spins becomes finite at the critical point in any tripartition of
mutually interacting spins described by the LMG model whereas it displays a
logarithmic divergence in a complementary bipartition.

From previous studies, a question is raised which correlations are
responsible for the divergent behavior of many-body systems at critical
points. To clarify this question, one need to distinguish classical
correlation from quantum one since the later one is much sensitive to the
disturbance of environments. By recognizing the discrepancy between quantum
extensions of two equivalent expressions for the classical mutual
information \cite{QD0}, Olliver and Zurek \cite{QD1,QD2} introduced quantum
discord (QD) that is sufficiently to qualify the total amount of quantum
correlation including entanglement in a composite system and classical
correlation (CC). In this paper, in order to analytically and clearly answer
the raised question, taking the LMG model as an example, we compute the QD
and the CC of two macroscopic sets of the mutually interacting spins in the
cases of both bipartition and tripartition. For completeness and comparison,
the EoF and the logarithmic negativity (LN) are also computed. We find that
at the critical point both the CC and quantum correlations, including QD,
EoF and LN, are always divergent in the bipartition. However, in the
tripartition the quantum correlations remain finite and the CC is still
divergent. The result may lead to a conjecture that the singular behaviour
of observables of a many-body system with finite temperatures at critical
points comes from the CC divergency.

This paper is organized as follows. In section 2, the model is introduced.
In section 3, correlations in a bipartition setting are studied and detailed
discussions are given. In section 4, correlations in a tripartition setting
are investigated. Finally, a brief summary is given in section 5.

\section{The Model}

The LMG model describes a collection of mutually interacting $N$ spins-$1/2$
on the $x$-$y$ plane with an external field applied along the $z$ direction.
The Hamiltonian of LMG model reads
\begin{equation}
H=-\frac{1}{N}\sum_{i(<j)=1}^{N}\left( \sigma _{i}^{x}\sigma _{j}^{x}+\gamma
\sigma _{i}^{y}\sigma _{j}^{y}\right) -h\sum_{i=1}^{N}\sigma _{i}^{z},
\label{LMG1}
\end{equation}%
where $\sigma _{i}^{\beta }$ $\left( \beta =x,y,z\right) $ are the Pauli
matrices for a spin at position $i$, $N$ is the total number of spins, $%
0\leq \gamma <1$ is an anisotropy parameter and $h$ is an external magnetic
field applied along the $z$ direction.

In terms of the total spin operators $S_{\beta }=\sum_{i}\sigma _{i}^{\beta
}/2$, the Hamiltonian (\ref{LMG1}) can be rewritten as%
\begin{equation}
H=-\frac{1}{N}\left( S_{x}^{2}+\gamma S_{y}^{2}\right) -hS_{z}.  \label{LMG2}
\end{equation}%
The ground-state properties of the LMG model have been found by use of a
mean-field approach \cite{LMG5, phase2}. The LMG model undergoes a
second-order phase transition at $h=1$. For $h>1$, the ground state is a
symmetrical and fully polarized state where all the spins are along the
external field direction. For $h<1$, the corresponding ground state is
two-fold degenerate \cite{LMG5, phase2}.

\section{The quantum and classical correlations in a bipartition}

In this section, we divide the $N$ spins into two groups and investigate the
ground-state quantum and classical correlations between the two spin groups.
To do so, we first need to determine the lowest energy state of the LMG
model for a given external field. Thus, a rotation transformation to the
total spin operators around the $y$ axis is introduced as follows
\begin{equation}
\left(
\begin{array}{c}
S_{x} \\
S_{y} \\
S_{z}%
\end{array}%
\right) =\left(
\begin{array}{ccc}
\cos \theta _{0} & 0 & \sin \theta _{0} \\
0 & 1 & 0 \\
-\sin \theta _{0} & 0 & \cos \theta _{0}%
\end{array}%
\right) \left(
\begin{array}{c}
\widetilde{S}_{x} \\
\widetilde{S}_{y} \\
\widetilde{S}_{z}%
\end{array}%
\right) .  \label{rotate}
\end{equation}%
In (\ref{rotate}), $\theta _{0}$ stands for the value of the rotation angle
which is chosen to make the expectation of the Hamiltonian (2) in the ground
state $\left\langle H\right\rangle $ be minimum. $\theta _{0}=0$ for the
symmetrical phase with $h>1$, and $\theta _{0}=\arccos h$ for the broken
phase with $0\leq h<1$ \cite{LMG5, phase2}. Substituting Eq. (\ref{rotate})
into (\ref{LMG2}), one obtains
\begin{eqnarray}
H &=&-\frac{1}{2N}\cos \theta _{0}\sin \theta _{0}\left( \widetilde{S}_{+}%
\widetilde{S}_{z}+\widetilde{S}_{-}\widetilde{S}_{z}+\widetilde{S}_{z}%
\widetilde{S}_{+}+\widetilde{S}_{z}\widetilde{S}_{-}\right)  \nonumber \\
&&-\frac{1}{4N}\left( \left( \cos ^{2}\theta _{0}-\gamma \right) \left(
\widetilde{S}_{+}^{2}+\widetilde{S}_{-}^{2}\right) +\left( \cos ^{2}\theta
_{0}+\gamma \right) \left( \widetilde{S}_{+}\widetilde{S}_{-}+\widetilde{S}%
_{-}\widetilde{S}_{+}\right) \right)  \nonumber \\
&&-\frac{1}{N}\sin ^{2}\theta _{0}\widetilde{S}_{z}^{2}+\frac{h\sin \theta
_{0}}{2}\left( \widetilde{S}_{+}+\widetilde{S}_{-}\right) -h\cos \theta _{0}%
\widetilde{S}_{z}.  \label{RT2}
\end{eqnarray}%
When working out Eq. (\ref{RT2}), we have used the relations $\widetilde{S}%
_{x}=\left( \widetilde{S}_{+}+\widetilde{S}_{-}\right) /2$ and $\widetilde{S}%
_{y}=\left( \widetilde{S}_{+}-\widetilde{S}_{-}\right) /\left( 2i\right) $.

We now split the $N$ spins into two groups and consequently write the total
spin operators as $\widetilde{S}_{\beta }=\widetilde{S}_{\beta }^{\left(
1\right) }+\widetilde{S}_{\beta }^{\left( 2\right) }$. In the
Holstein-Primakoff representation \cite{LMG5}, the spin operators $%
\widetilde{S}_{\beta }^{\left( k\right) }(k=1,2)$ for each of the spin
groups can be written as
\begin{equation}
\widetilde{S}_{z}^{\left( k\right) }=N_{k}/2-a_{k}^{\dagger }a_{k},
\label{HP1}
\end{equation}%
\begin{equation}
\widetilde{S}_{+}^{\left( k\right) }=\left( N_{k}-a_{k}^{\dagger
}a_{k}\right) ^{1/2}a_{k},  \label{HP2}
\end{equation}%
\begin{equation}
\widetilde{S}_{-}^{\left( k\right) }=a_{k}^{\dagger }\left(
N_{k}-a_{k}^{\dagger }a_{k}\right) ^{1/2},  \label{HP3}
\end{equation}%
where $a_{k}$ and $a_{k}^{\dagger }$ are bosonic annihilation and create
operators and $N_{k}$ denotes the spin number in the $k$th group under the
condition $N=N_{1}+N_{2}$.

Upon substituting Eqs. (\ref{HP1})-(\ref{HP3}) into (\ref{RT2}) and
expanding $H$ as a series of powers $1/N_{k}$, and keeping the lowest order,
one obtains%
\begin{eqnarray}
H &=&-\frac{m^{2}-\gamma }{4N}\left( N_{1}a_{1}^{2}+\sqrt{N_{1}N_{2}}%
a_{1}a_{2}+\sqrt{N_{1}N_{2}}a_{2}a_{1}+N_{2}a_{2}^{2}+h.c.\right)  \nonumber
\\
&&-\frac{m^{2}+\gamma }{4N}\left( N_{1}a_{1}a_{1}^{\dagger }+\sqrt{N_{1}N_{2}%
}a_{1}a_{2}^{\dagger }+\sqrt{N_{1}N_{2}}a_{2}a_{1}^{\dagger
}+N_{2}a_{2}a_{2}^{\dagger }+h.c.\right)  \nonumber \\
&&+\left( 1-m^{2}+hm\right) \left( a_{1}^{\dagger }a_{1}+a_{2}^{\dagger
}a_{2}\right) +Const,  \label{HPH}
\end{eqnarray}%
with $m=\cos \theta _{0}$. The Hamiltonian (\ref{HPH}) is of a quadratic
form of the bosonic annihilation and creation operators. It can be
diagonalized by introducing the Bogoliubov transformation
\begin{equation}
a_{1}=\left( \cosh \frac{\Theta }{2}b_{1}+\sinh \frac{\Theta }{2}%
b_{1}^{\dagger }\right) \sqrt{\tau _{1}}+b_{2}\sqrt{\tau _{2}},  \label{BT1}
\end{equation}%
\begin{equation}
a_{2}=\left( \cosh \frac{\Theta }{2}b_{1}+\sinh \frac{\Theta }{2}%
b_{1}^{\dagger }\right) \sqrt{\tau _{2}}-b_{2}\sqrt{\tau _{1}},  \label{BT2}
\end{equation}%
where $b_{i}(i=1,2)$ are new bosonic operators and $\tau _{k}=N_{k}/N$ with $%
\sum_{k}\tau _{k}=1$. If choosing
\begin{equation}
\tanh \Theta =-s/r,  \label{tanh}
\end{equation}%
with $s=\gamma -m^{2},$ and $r=2hm-3m^{2}+2-\gamma ,$ the Hamiltonian (\ref%
{HPH}) can be written in the diagonal form except an irrelevant constant to
the present investigation
\begin{equation}
H=\Delta _{1}b_{1}^{\dagger }b_{1}+\Delta _{2}b_{2}^{\dagger }b_{2},
\label{BTH}
\end{equation}%
where%
\begin{equation}
\Delta _{1}=\frac{1}{2}\left[ \left( 2hm-3m^{2}+2-\gamma \right) \cosh
\Theta -\left( m^{2}-\gamma \right) \sinh \Theta \right] ,  \label{gap1}
\end{equation}%
\begin{equation}
\Delta _{2}=\frac{1}{2}\left( 2hm-3m^{2}+2-\gamma \right) .  \label{gap2}
\end{equation}

The Hamiltonian (\ref{BTH}) represents two independent harmonic oscillators
which ground state $\left\vert \psi _{0}\right\rangle $ is a Gaussian state,
defined as $b_{i}\left\vert \psi _{0}\right\rangle =0$. Therefore, the
ground state of the LMG model can be fully characterized by the covariance
matrix with the elements $\Gamma _{ij}=\left\langle \psi _{0}\right\vert
\left\{ \widehat{R}_{i},\widehat{R}_{j}\right\} \left\vert \psi
_{0}\right\rangle $, where $\widehat{\mathbf{R}}=\left( \widehat{x}_{1},%
\widehat{p}_{1},\widehat{x}_{2},\widehat{p}_{2}\right) $ with canonical
coordinates $\widehat{x}_{k}=\left( a_{k}^{\dagger }+a_{k}\right) /\sqrt{2}$
and momenta $\widehat{p}_{k}=i\left( a_{k}^{\dagger }-a_{k}\right) /\sqrt{2}$%
. By use of the Bogoliubov transformation (\ref{BT1}) and (\ref{BT2}), one
can obtain the explicit expression for the covariance matrix
\begin{equation}
\Gamma =\left(
\begin{array}{cccc}
A_{1}\tau _{1}+1 & 0 & A_{1}\sqrt{\tau _{1}\tau _{2}} & 0 \\
0 & B_{1}\tau _{1}+1 & 0 & B_{1}\sqrt{\tau _{1}\tau _{2}} \\
A_{1}\sqrt{\tau _{1}\tau _{2}} & 0 & A_{1}\tau _{2}+1 & 0 \\
0 & B_{1}\sqrt{\tau _{1}\tau _{2}} & 0 & B_{1}\tau _{2}+1%
\end{array}%
\right) =\left(
\begin{array}{cc}
G_{1} & C_{1} \\
C_{1} & G_{2}%
\end{array}%
\right) ,  \label{CM1}
\end{equation}%
where%
\begin{equation}
A_{1}=\sqrt{\left( r-s\right) /\left( r+s\right) }-1,  \label{CME1}
\end{equation}%
\begin{equation}
B_{1}=\sqrt{\left( r+s\right) /\left( r-s\right) }-1,  \label{CME2}
\end{equation}%
and $G_{i},C_{i}$ are $2\times 2$ matrices. For the simplicity of
expressions in the following, we set $\sqrt{\left( r+s\right) /\left(
r-s\right) }=\alpha $. In the symmetrical $\left( h\geq 1\right) $\ and
broken $\left( 0\leq h<1\right) $ phases the parameter $\alpha $ reads%
\begin{equation}
\alpha =\left\{
\begin{array}{lc}
\sqrt{\left( h-1\right) /\left( h-\gamma \right) }, & h\geq 1 \\
\sqrt{\left( 1-h^{2}\right) /\left( 1-\gamma \right) }, & 0\leq h<1%
\end{array}%
\right.  \label{prefactor}
\end{equation}%
By performing a like-Bogoliubov transformation \cite{Duan}, the covariance
matrix (\ref{CM1}) can be written in the standard form
\begin{equation}
\Gamma _{sf}=\left(
\begin{array}{cccc}
a & 0 & c_{1} & 0 \\
0 & a & 0 & c_{2} \\
c_{1} & 0 & b & 0 \\
0 & c_{2} & 0 & b%
\end{array}%
\right) ,  \label{CM2}
\end{equation}%
where $a,b,c_{1}$ and $c_{2}$ are determined by
\begin{equation}
a^{2}=\det G_{1}=A,  \label{A}
\end{equation}%
\begin{equation}
b^{2}=\det G_{2}=B,  \label{B}
\end{equation}%
\begin{equation}
c_{1}c_{2}=\det C_{1}=C,  \label{C}
\end{equation}%
\begin{equation}
\left( ab-c_{1}^{2}\right) \left( ab-c_{2}^{2}\right) =\det \Gamma =D,
\label{D}
\end{equation}%
with%
\begin{equation}
A=B=\alpha ^{-1}\left[ \alpha \tau _{1}+\left( 1-\tau _{1}\right) \right]
\left( \tau _{1}+\alpha \left( 1-\tau _{1}\right) \right) ,  \label{AB}
\end{equation}%
\begin{equation}
C=\left[ 2-\alpha -\alpha ^{-1}\right] \left[ 1-\tau _{1}\right] \tau _{1},
\label{ABC1}
\end{equation}%
and $D=1$.

Correspondingly, the simplectic eigenvalues of the covariance matrix (\ref%
{CM2}) are given by
\begin{equation}
\nu _{\pm }^{2}=\left( M\pm \sqrt{M^{2}-4D}\right) /2,  \label{SE1}
\end{equation}%
with $M=A+B+2C$.

According to Ref. \cite{Gaussian discord}, the classical and quantum
correlations of the Gaussian state characterized by the covariance matrix (%
\ref{CM2}) can be respectively calculated by the formula
\begin{equation}
CC=f\left( \sqrt{A}\right) -f\left( \sqrt{E^{\min }}\right) ,  \label{CC1}
\end{equation}%
\begin{equation}
QD=f\left( \sqrt{B}\right) -f\left( \nu _{-}\right) -f\left( \nu _{+}\right)
+f\left( \sqrt{E^{\min }}\right) ,  \label{QD1}
\end{equation}%
where $E^{\min }=\left( 2C^{2}+\left( B-1\right) \left( D-A\right)
+2\left\vert C\right\vert \sqrt{C^{2}+\left( B-1\right) \left( D-A\right) }%
\right) /\left( B-1\right) ^{2}$ for $\left( D-AB\right) ^{2}\leq \left(
1+B\right) C^{2}\left( A+D\right) $, and when it comes to other cases $%
E^{\min }$ is determined by $\left( AB-C^{2}+D-\sqrt{C^{4}+\left(
D-AB\right) ^{2}-2C^{2}\left( AB+D\right) }\right) /\left( 2B\right) $, and $%
f\left( x\right) =\left( \frac{1+x}{2}\right) \ln \left( \frac{1+x}{2}%
\right) -\left( \frac{x-1}{2}\right) \ln \left( \frac{x-1}{2}\right)$.

Upon substituting Eqs. (\ref{AB})-(\ref{SE1}) into Eqs. (\ref{CC1}) and (\ref%
{QD1}), the explicit expressions for the CC and the QD can be obtained as
\begin{equation}
CC=QD=\frac{\sqrt{A}+1}{2}\ln \frac{\sqrt{A}+1}{2}-\frac{\sqrt{A}-1}{2}\ln
\frac{\sqrt{A}-1}{2}.  \label{QD2}
\end{equation}%
It is noted that the expression of the CC and the QD are the same as that of
the entanglement entropy obtained in Ref. \cite{LMG4}.

From (\ref{CM2}), we can also obtain the entanglement of formation (EoF)
between the two divided spin groups \cite{6}
\begin{equation}
EoF=f\left( \Delta \right) ,  \label{EOF}
\end{equation}%
where
\begin{equation}
f\left( \Delta \right) =c_{+}\left( \Delta \right) \log _{2}\left(
c_{+}\left( \Delta \right) \right) -c_{-}\left( \Delta \right) \log
_{2}\left( c_{-}\left( \Delta \right) \right) ,  \label{EOF1}
\end{equation}%
\begin{equation}
c_{\pm }\left( \Delta \right) =\left( \Delta ^{-1/2}\pm \Delta ^{1/2}\right)
^{2}/4,  \label{EOF2}
\end{equation}%
\begin{equation}
\Delta =a-c,  \label{EOF3}
\end{equation}%
\begin{equation}
c=\alpha ^{-1/2}\sqrt{(-1+\alpha )^{2}(1-\tau _{1})\tau _{1}}.  \label{EOF4}
\end{equation}

From Eq. (\ref{CM2}), one can also calculate the logarithmic negativity (LN)
\cite{7} which has been obtained by Wichterich et al in Ref. \cite{LMG14}.

In Fig. 1, the various correlations such as CC, QD, EoF and LN are plotted
against the external field $h$. It is clearly shown that all the
correlations between the two spin groups diverge at the critical point $h=1$%
. By comparing the two figures, one may find that the anisotropic parameter $%
\gamma $ has little impact on the singular behaviour of these correlations
and the CC, QD, EoF and LN display the same divergency at the critical point
although they describe the different correlations. In Fig. 2, the
correlations versus the external field are shown for different divisions of
bipartition. It is observed that the divergent behavior of the correlations
at the critical point is hardly affected by the bipartition way.

In order to analytically investigate the critical behavior of the CC, the QD
and the EoF, we expand Eqs. (\ref{QD2}) and (\ref{EOF}) at the critical
point ($h=1$) and obtain
\begin{equation}
CC=QD=-\frac{1}{4}\ln \left( h-1\right) +\frac{1}{4}\ln \left( 1-\gamma
\right) +\frac{1}{2}\ln \tau _{1}\left( 1-\tau _{1}\right) -\ln 2.
\label{Exp1}
\end{equation}%
and
\begin{equation}
EoF=-\frac{1}{4}\log _{2}\left( h-1\right) +\frac{1}{4}\log _{2}\left(
1-\gamma \right) +\frac{1}{2}\log _{2}\tau _{1}\left( 1-\tau _{1}\right) -1.
\label{EoF5}
\end{equation}%
Eq. (\ref{Exp1}) shows that when reaching the critical point the CC and QD
diverge as $-\frac{1}{4}\ln \left( h-1\right) $ which is consistent with
that appears in Figs. 1 and 2. Interestingly, the singular behavior of the
CC and QD is really the same as that of the logarithmic negativity \cite%
{LMG14}, the entanglement entropy \cite{LMG4} and the single-copy
entanglement \cite{LMG9}. From Eq. (\ref{EoF5}) we know that the EoF
diverges as $-\frac{1}{4}\log _{2}\left( h-1\right) $ at the critical point
and behaves slightly different from the CC and the QD.

Based on the scaling hypothesis proposed in Refs. \cite{LMG4,LMG1}, the
finite-size scaling behavior of the CC and QD can be straightforwardly
extracted from Eq. (\ref{Exp1})
\begin{equation}
CC=QD\sim \frac{1}{6}\ln N+\frac{1}{6}\ln \left( 1-\gamma \right) +\frac{1}{2%
}\ln \tau _{1}\left( 1-\tau _{1}\right) .  \label{scale}
\end{equation}%
This finite-size scaling behavior is identical to that of the logarithmic
negativity \cite{LMG14}, the entanglement entropy \cite{LMG4}, the geometric
entanglement and the single-copy entanglement \cite{LMG9}. Therefore, all
the correlations between artificial divided two parts of the mutually
interacting spins in the LMG model obey the same critical scaling law.

\section{The classical and quantum correlations in a tripartition}

In this section, we divide the mutually interacting $N$ spins in the LMG
model into three groups, each of which has $N_{i}$ spins under the condition
$N=N_{1}+N_{2}+N_{3}$, and investigate correlations between any two groups.
In this case, if we consider the first and third groups, we need to trace
the variable of the second group. Thus, the spins in the second group plays
a role of the environment to the spins in the first and third groups, and
the spins in the groups under consideration is generally in a mixed state.

Following the same procedure as shown in the preceding section, we can
diagonalize the Hamiltonian (\ref{LMG1}) and obtain the ground state of the
LMG model, from which the density matrix of the ground state can be built.
By tracing the density matrix over the variable of spins in the second
group, one can obtain the reduced density matrix for spins in the first and
third groups. It is obvious that the reduced density matrix is also of a
Gaussian state. The covariance matrix of the reduced density matrix is found
to be \cite{LMG14}

\begin{equation}
\Gamma =\left(
\begin{array}{cccc}
A_{1}\tau _{1}+1 & 0 & A_{1}\sqrt{\tau _{1}\tau _{3}} & 0 \\
0 & B_{1}\tau _{1}+1 & 0 & B_{1}\sqrt{\tau _{1}\tau _{3}} \\
A_{1}\sqrt{\tau _{1}\tau _{3}} & 0 & A_{1}\tau _{3}+1 & 0 \\
0 & B_{1}\sqrt{\tau _{1}\tau _{3}} & 0 & B_{1}\tau _{3}+1%
\end{array}%
\right) .  \label{CM3}
\end{equation}%
If one sets $\tau _{1}=\tau _{3}=\tau <1/2$, the standard form of (\ref{CM3}%
) is the same as (\ref{CM2}) which elements are determined by
\begin{equation}
A=B=\alpha ^{-1}\left( \alpha \tau +\left( 1-\tau \right) \right) \left(
\tau +\alpha \left( 1-\tau \right) \right) ,  \label{AB2}
\end{equation}%
\begin{equation}
C=-\alpha ^{-1}\left( \alpha -1\right) ^{2}\tau ^{2},  \label{AB3}
\end{equation}%
\begin{equation}
D=\alpha ^{-1}\left( \alpha +2\left( \alpha -1\right) ^{2}\left( \tau -2\tau
^{2}\right) \right)  \label{AB4}
\end{equation}%
according to Eqs. (20)-(23). In this case, the reduced density matrix is of
a symmetrical Gaussian state \cite{Duan}. The symplectic eigenvalues of Eq. (%
\ref{CM3}) are found to be
\begin{equation}
\begin{array}{c}
\nu _{-}=1, \\
\nu _{+}=\alpha ^{-1/2}\sqrt{\alpha +2\left( \alpha -1\right) ^{2}\left(
\tau -2\tau ^{2}\right) }%
\end{array}
\label{SE22}
\end{equation}

Upon substituting Eqs. (\ref{AB2})-(\ref{SE22}) into Eqs. (\ref{CC1})-(\ref%
{QD1}), one can work out the the CC and the QD between spins in the first
and third groups with $E^{\min }=1$ for $h=\sqrt{\gamma }$, and $E^{\min
}=\left( -2\alpha ^{2}-4da\tau -d\left( 1+\left( \alpha -8\right) \alpha
\right) \tau ^{2}+2d^{2}\tau ^{3}+\left\vert d^{3/2}\right\vert \left(
\alpha +1\right) \tau ^{2}\left( 1-2\tau \right) \right) /\mu $ for other
circumstance with $d=\left( \alpha -1\right) ^{2}$ and $\mu =2\alpha \left(
\alpha \left( \tau -1\right) -\tau \right) \left( 1+\left( \alpha -1\right)
\tau \right) $. Since the analytical expressions for the CC and the QD are
much lengthy, we here have to give up to explicitly write them out.

The entanglement of formation (EoF) can be obtained from Eq. (\ref{EOF})
with
\begin{equation}
\Delta =\sqrt{\left( \sqrt{A}-k_{1}\right) \left( \sqrt{A}-k_{2}\right) },
\label{eof2}
\end{equation}%
where%
\begin{equation}
k_{1}=\sqrt{\frac{\left( \alpha -1\right) ^{2}\tau ^{2}\left( 1+\left(
\alpha -1\right) \tau \right) }{\alpha \left( \alpha +\tau -\alpha \tau
\right) }},  \label{eof3}
\end{equation}%
\begin{equation}
k_{2}=\frac{\left( \alpha -1\right) ^{2}\tau ^{2}}{\alpha \ast k_{1}}.
\label{eof4}
\end{equation}

In Fig. 3, the various correlations for an equal tripartition $\tau
_{1}=\tau _{3}=1/3$ are plotted as a function of the magnetic field $h$. It
is clearly observed that the CC diverges whereas all the quantum correlation
measures such as QD, EoF and LN remain finite at the critical point.

To clearly look into the behavior of the CC and QD at the critical point, we
expand the analytical expression of the QD at $h=1$ and obtain
\begin{equation}
QD=\ln \frac{\sqrt{1-\tau }}{2\sqrt{2}}+\frac{1}{2}\ln \left( \frac{\sqrt{%
2\left( 1-\tau \right) }+1}{\sqrt{2\left( 1-\tau \right) }-1}\right) ^{\sqrt{%
2\left( 1-\tau \right) }}.  \label{ExpC}
\end{equation}%
Eq. (\ref{ExpC}) shows that the QD indeed remains finite at the critical
point. Moreover, the value of the QD is irrelative to the anisotropy
parameter $\gamma $. This universal character is much similar to that of the
logarithmic negativity as found in Ref. \cite{LMG14}. Eq. (\ref{ExpC}) also
shows that when $\tau$ approaches to 1/2, QD diverges as that in the
bipartition setting.

In the similar way, we can find the analytical expression for the classical
correlation of the LMG model around the critical point
\begin{eqnarray}
CC &=&-\frac{1}{4}\ln \left( h-1\right) +\frac{1}{4}\ln \left( 1-\gamma
\right)  \nonumber \\
&&+\frac{1}{2}\ln \left\{ \frac{\tau \left( 1-\tau \right) }{\left( 1-2\tau
\right) }\left( \frac{\sqrt{2\left( 1-\tau \right) }-1}{\sqrt{2\left( 1-\tau
\right) }+1}\right) ^{\sqrt{2\left( 1-\tau \right) }}\right\} .  \label{Exp2}
\end{eqnarray}%
In contrast to the QD, the classical correlation between the two spin groups
diverges as $-\frac{1}{4}\ln \left( h-1\right) $ at the critical point. This
divergent behavior is the same as that obtained in the bipartition setting.

In quantum information theory, the total correlation of a bipartite quantum
system is measured by the mutual information \cite{total1,total2}.
Qualitatively, the total correlation equals to the QD plus the CC. From the
present result, it is very clear that in a tripartite setting the classical
correlation is responsible for the divergency of the total correlations at
the critical point \cite{LMG14}.

The critical behaviour of EoF can also be investigated from Eqs.(\ref{EOF}),
(\ref{eof2})-(\ref{eof4}). However, the analytical expression of it is too
lengthy to be explicitly written here. We just give the numerical results.
In Fig. 4, the EoF and QD are plotted as a function of the partition
parameter $\tau $ at the critical point $h=1$. It clearly shows that when $%
\tau $ is less than $1/2$ the EoF and QD remain finite. When $\tau $ reaches
$1/2$ and the tripartition reduces to the bipartition, the EoF and QD go
from finite to infinity.

\section{Summary}

The Lipkin-Meshkov-Glick (LMG) model describes a collection of mutually
interacting spins-1/2 in an external magnetic field. By dividing spins of
the LMG model into two or three parts, we study the classical correlation
(CC) and quantum correlation measures such as the quantum discord (QD), the
entanglement of formation (EoF) and the logarithmic negativity (LN) between
the two spin groups. In the case of bipartition, where the two spin groups
are complementary and their ground state must be of a pure state, we find
that the classical correlations and all the quantum correlations are
divergent in the same singular behaviour at the critical point of the LMG
model. In the case of tripartition, however, the classical correlation is
still divergent but all the quantum correlation measures remain finite at
the critical point. In a tripartition, the spin group traced out plays a
role of environments and the other two spin groups are general in a mixed
state. The present result shows that the classical correlation is very
robust but the quantum correlation is much frangible to the environment
disturbance. In the real situation, a many-body quantum system is
unavoidably to be coupled to its surroundings and is in a mixed state.
Therefore, the present result may lead to the conjecture that the classical
correlation is responsible for the singularity behaviour of physics
quantities at critical points of a many-body quantum system.

\section*{Acknowledgments}

This work was supported by the National Basic Research Program of China
(Grant No. 2010CB923102), Special Prophase Project on the National Basic
Research Program of China (Grant No. 2011CB311807), and the National Nature
Science Foundation of China (Grand No. 11074199).

\newpage

{\Large Figure Captions}

\bigskip

\bigskip

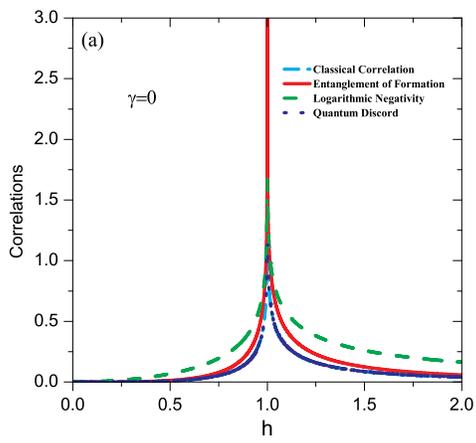
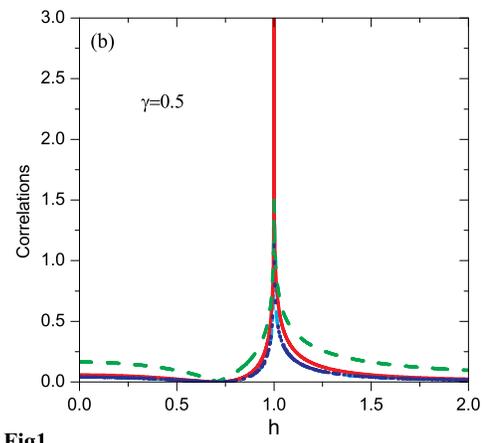
Fig. 1 \ Various correlations as a function of the magnetic field
$h$ for the bipartition with $\protect\tau _{1}=1/3$. The symbols
shown in the inset of Fig. 1(a) are applicable to the curves of
Fig. 1(b).

\bigskip

\bigskip

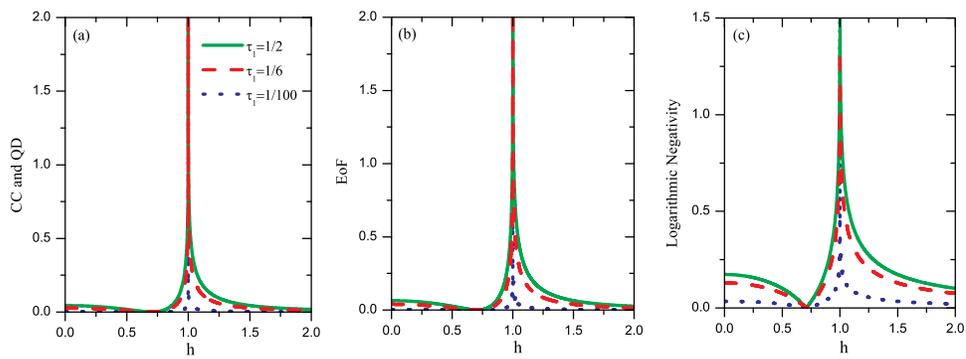
Fig. 2 \ Various correlations as function of the magnetic field
$h$ for the different divisions of bipartition with $\protect\tau
_{1}=1/2,1/6,1/100$ and $\protect\gamma =0.5$. The symbols shown
in the inset of Fig. 2(a) are applicable to the curves of Fig.
2(b) and Fig. 2(c).

\bigskip

\bigskip

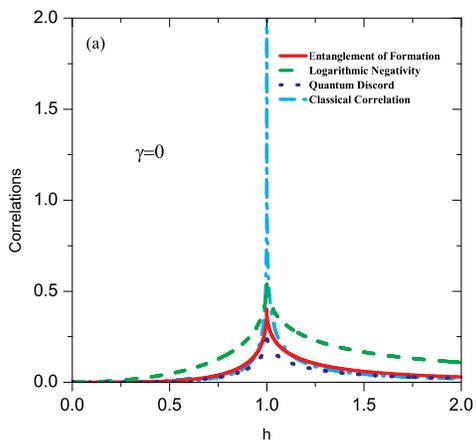
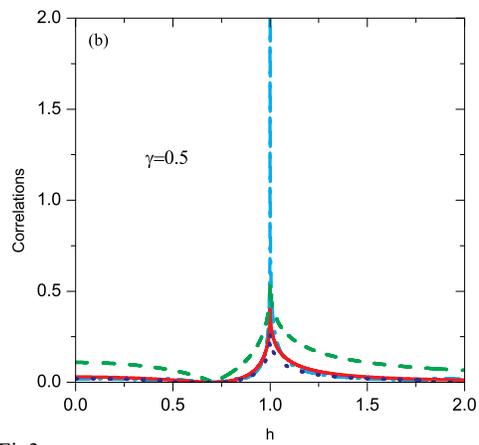
Fig. 3 \ The various correlations as function of the magnetic
field $h$ for an equal tripartition $\protect\tau
_{1}=\protect\tau _{2}=\protect\tau_{3}=1/3$. The symbols shown in
the inset of Fig. 3(a) are applicable to the curves of Fig. 3(b).

\bigskip

\bigskip

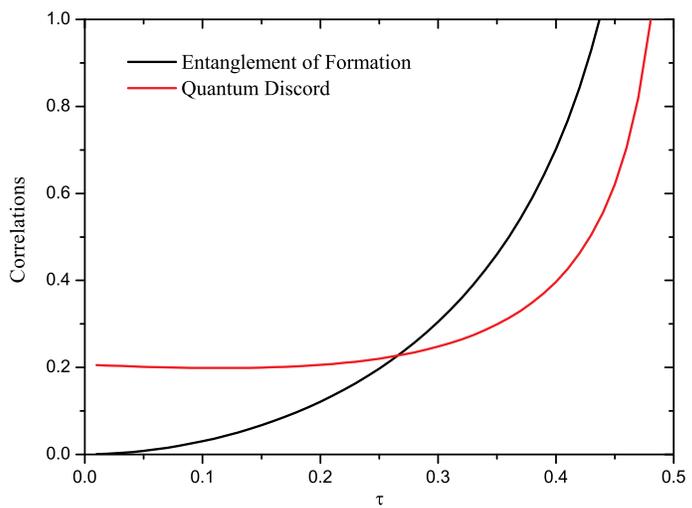
Fig. 4 \ The EoF and QD versus the partition parameter
$\protect\tau$ at the critical point $h=1$.


\section*{References}

\begin{thebibliography}{99}
\bibitem{1} Sachdev S 1999 \textit{Quantum Phase Transition} (England:
Cambridge)

\bibitem{2} Schr\"{o}dinger E 1935 Naturwiss. \textbf{23} 807; Schr\"{o}%
dinger E 1935 Naturwiss. \textbf{23} 823; E. Schr\"{o}dinger 1935 Naturwiss.
\textbf{23} 844

\bibitem{3} Einstein A, Podolski L, and Rosen N 1935 Phys. Rev. \textbf{47}
777

\bibitem{4} Vedral V, Plenio M B, Rippin M A, and Knight P L 1998 Phys. Rev.
Lett. \textbf{78} 2275

\bibitem{5} Wootters W K 1998 Phys. Rev. Lett. \textbf{80} 2245

\bibitem{6} Giedke G, Wolf M M, Kr\"{u}ger O, Werner R F, and Cirac J I 2003
Phys. Rev. Lett. \textbf{91} 107901

\bibitem{7} Vidal G and Werner R F 2002 Phys. Rev. A \textbf{65} 032314

\bibitem{10} Lambert N, Emary C, and Brandes T 2004 Phys. Rev. Lett. \textbf{%
92} 073602

\bibitem{11} Amico L, Fazio R, Osterloh A, and Vedral V 2008 Rev. Mod. Phys.
\textbf{80} 517

\bibitem{LMG4} Barthel T, Dusuel S, and Vidal J 2006 Phys. Rev. Lett.
\textbf{97} 220402

\bibitem{chain1} Osborne T J and Nielsen M A 2002 Phys. Rev. A \textbf{66}
032110

\bibitem{chain2} Vidal G, Latorre J I, Rico E and Kitaev A 2003 Phys. Rev.
Lett \textbf{90} 227902

\bibitem{13} Werlang T, Trippe C, Ribeiro G A P, and Rigolin G 2010 Phys.
Rev. Lett. \textbf{105} 095702

\bibitem{QD1} Ollivier H and Zurek W H 2001 Phys. Rev. Lett. \textbf{88}
017901

\bibitem{QD2} Zurek W H 2000 Ann. Phys (Berlin) \textbf{9} 855

\bibitem{LMG1} Dusuel S, Vidal J 2004 Phys. Rev. Lett \textbf{93} 237204

\bibitem{LMG2} Unanyan R G, Ionescu C, and Fleischhauer M 2005 Phys. Rev. A
\textbf{72} 022326

\bibitem{LMG3} Latorre J I, Or\'{u}s R, Rico E, and Vidal J 2005 Phys. Rev.
A \textbf{71} 064101

\bibitem{LMG5} Dusuel S, Vidal J 2005 Phys. Rev. B \textbf{71} 224420

\bibitem{LMG6} Ribeiro P, Vidal J, and Mosseri R 2007 Phys. Rev. Lett.
\textbf{99} 050402

\bibitem{LMG7} Morrison S and Parkins A S 2008 Phys. Rev. Lett. \textbf{100}
040403

\bibitem{LMG8} Kwok H M, Ning W Q, Gu S J and Lin H Q 2008 Phys. Rev. E
\textbf{78} 032103

\bibitem{LMG9} Or\'{u}s R, Dusuel S and Vidal J 2008 Phys. Rev. Lett.
\textbf{101} 025701

\bibitem{LMG10} Morrison S and Parkins A S 2008 Phys. Rev. A \textbf{77}
043810

\bibitem{LMG11} Ribeiro P, Vidal J and Mosseri R 2008 Phys. Rev. E \textbf{78%
} 021106

\bibitem{LMG13} Filippone M, Dusuel S, and Vidal J, 2011 Phys. Rev. A
\textbf{83} 022327

\bibitem{LMG14} Wichterich H, Vidal J, and Bose S 2010 Phys. Rev. A \textbf{%
81} 032311

\bibitem{QD0} Henderson L and Vedral V 2001 J. Phys. A \textbf{34} 68899

\bibitem{phase2} Botet R and Jullien R 1983 Phys. Rev. B \textbf{28} 3955

\bibitem{Duan} Duan L M, Giedke G, Cirac J I, and Zoller P 2000 Phys. Rev.
Lett. \textbf{84} 2722

\bibitem{Gaussian discord} Adesso G, Datta A 2010 Phys. Rev. Lett. \textbf{%
105} 030501

\bibitem{total1} Groisman B, Popescu S, and Winter A 2005 Phys. Rev. A
\textbf{72} 032317

\bibitem{total2} Schumacher B, Westmoreland M D 2006 Phys. Rev. A \textbf{74}
042305
\end{thebibliography}
\end{document}